\documentclass[pra,twocolumn,superscriptaddress]{revtex4-2}
\usepackage{multirow,eurosym,amssymb,amsfonts,setspace,graphicx,bm,float,amssymb}

\usepackage{color}

\newcommand{\bra}[1]{\langle #1|}
\newcommand{\ket}[1]{|#1\rangle}

\def\be{\begin{equation}}
\def\ee{\end{equation}}

\def\bsplit{\begin{split}}
\def\nsplit{\end{split}}

\begin{document}
\title{Relativistic Quantum Simulation under Periodic and Dirichlet Boundary Conditions: A First-Quantised Framework for Near-Term Devices}

\date{\today}

\author{Jaewoo Joo} 
\email{\href{mailto:jaewoojoo@port.ac.uk}{jaewoojoo@port.ac.uk} }
\address{School of Mathematics and Physics, University of Portsmouth, Portsmouth PO1 3QL, UK}

\author{Timothy P. Spiller} 
\address{York Centre for Quantum Technologies, School of Physics, Engineering and Technology, University of York, York, YO10 5DD, UK}

\author{Kyunghyun Baek}
\address{Institute for Convergence Research and Education in Advanced Technology, Yonsei University, Seoul 03722, Republic of Korea}

\author{Jeongho Bang}
\address{Institute for Convergence Research and Education in Advanced Technology, Yonsei University, Seoul 03722, Republic of Korea}

\begin{abstract} 

We present a new recipe for relativistic quantum simulation using the first quantisation approach, under periodic (PBC) and Dirichlet (DBC) boundary conditions. The wavefunction is discretised across a finite grid represented by system qubits, and the squared momentum operator is expressed using the finite-difference method based on quantum translation operations. The relativistic kinetic energy is approximated through a perturbative expansion of the total kinetic Hamiltonian, incorporating higher-order momentum terms. 
The approach would allow variational optimisation of appropriate ansatz states to estimate both non-relativistic and relativistic ground-state energies on a quantum computer. This work offers a practical route to simulating relativistic effects on near-term quantum devices, supporting future developments in quantum physics and chemistry.

\end{abstract}

\maketitle 

\section{Introduction}

Simulating quantum systems is crucial for understanding the intrinsic nature of quantum phenomena, but remains a challenging task in practice \cite{Feynmann}. Due to the principles of superposition and entanglement, the complexity of describing a quantum system grows exponentially with its system size, making it difficult to compute the desired results for large quantum systems. In addition, the relativistic effects of quantum physics must be incorporated to accurately describe many important quantum phenomena. One of the standard approaches to studying relativistic quantum effects is to use perturbation theory in a quantum system \cite{Kais2023}, which accumulates quantum relativistic effects beyond the non-relativistic framework.

We present a new quantum simulation method for determining a relativistic ground-state energy, for the cases of two different boundary conditions. In a chosen frame, the relativistic time-independent single-particle Hamiltonian is 
\begin{eqnarray}
\hat{H}^{tot}_{re} (\vec{r}) &=&  \hat{H}^{K}_{re}(\vec{r}) +  \hat{V} (\vec{r})\,, \label{TotalH01}
\end{eqnarray}
where $\hat{H}^{K}_{re} $ and $\hat{V} $ are the relativistic kinetic and potential Hamiltonian components, dependent on the particle's position $\vec{r}$ in that frame. 
From special relativity, the kinetic part of the Hamiltonian with positive energy is represented in the form of a square root. Therefore, it can be expanded to generate higher-order correction terms dependent on mass $m$ and light speed $c$, in a valid expansion domain. For example, in the regime where relativistic effects are just becoming relevant, $\left< \hat{p}^2 \right> \ll m^2\,c^2$, the perturbation terms \cite{Edery2018, KG} are given by 
\begin{eqnarray}
\hat{H}^{K}_{re} &=& m\,c^2 \left( \sqrt{1+ { \hat{p}^2 \over m^2 \, c^2}} - 1\right) = m\,c^2 \sum_{l =1}^{\infty} { \cal G } (\hat{p}^{2l}) \,,~~\nonumber \\
&=& {1 \over 2 m} \hat{p}^2 - {1 \over 8 m^3 \, c^2}  \hat{p}^4 + {1 \over 16 m^5 \, c^4}  \hat{p}^6 + ... \,,
\label{Normal_HK01}
\end{eqnarray}
where ${ \cal G }$ is a function of the squared momentum operator $\hat{p}^{2}$. Clearly in the ultra relativistic limit $\left< \hat{p}^2 \right> \gg m^2\,c^2$, the expansion would be of a different form.

In this work we focus on the regime where corrections beyond the standard non-relativistic approach are relevant, considering a quantum system in a one-dimensional (1D) configuration, with the continuous-variable 1D momentum operator $\hat{p}$ given by the derivative operator $-i \hbar \,{d / d x}$. For an eigen-state of the total Hamiltonian $\ket{\Psi_n (x)}$ with non-negative quantum number $n$, the expectation value of the relativistic kinetic energy is given by \begin{eqnarray}
{\cal E}^{K}_n && = \bra{\Psi_n (x)} \hat{H}^{K}_{re} (x) \ket{\Psi_n (x)} , \nonumber \\
&&= - m c^2\left( \sum_{l=1}^{\infty} \alpha_l \, \left( \lambdabar_m\right)^{2l} \, \Big< \Psi_n (x) \Big| {d^{\,2l} \over d x^{2l}} \Big| \Psi_n (x) \Big> \right) .~~~ \label{Kin-Energy01} 
\end{eqnarray}
The coefficients $\alpha_l$ and the reduced Compton wavelength $\lambdabar_m$ originate from the Klein-Gordon equation \cite{KG},
\begin{eqnarray}
\alpha_l &=& \frac{{\cal C}^{2l}_{l} }{(2l-1) \, 4^l }  \\
{ \lambdabar_m } &=&  { \hbar \over m c}\, ,
\end{eqnarray}
with ${\cal C}^{2l}_{l} = \left(
\begin{array}{c}
2l \\
l \\
\end{array}
\right) = (2l)! / (l!)^2$.
Since the potential energy is expressed by the expectation value of the 1D potential operator ${\cal E}^{V}_n = \bra{\Psi_n (x)} \hat{V}(x) \ket{\Psi_n (x)}$,
the total energy becomes
\begin{eqnarray}
&& {\cal E}^{tot}_n =  {\cal E}^{K}_n +  {\cal E}^{V}_n \, .
\end{eqnarray}
For example, the ground-state energy (lowest energy) is given by ${\cal E}^{tot}_0 = {\cal E}^{K}_0 +  {\cal E}^{V}_0$ for the case $n=0$.

To estimate the total energy, we could utilise either first or second quantisation methods, which offer two fundamentally different frameworks for quantum systems. The second quantisation method is used widely for conventional simulations of quantum systems \cite{KG}, supporting much of the recent progress with quantum simulations for quantum chemistry and/or many-body physics \cite{Many-body, Chemistry}. Alternatively, the first quantisation method often provides a more intuitive understanding, with benefits for complex quantum systems with non-linearities (e.g., quantum simulation for Bose-Einstein condensates) \cite{Simon2023,PRX2021,JJ2020}.

Our approach is a first quantised method, designed to estimate ground-state energies. We assume that a discretised quantum state in position space is represented by $\ket{\psi (x)} = \sum_{j=0}^{2^L-1} c_{j} \ket{j}$, with $L$ qubits providing $2^L$ grid points spanning the dimensionless position range $0\le x < 1$ to describe a 1D spatial wavefunction \cite{JJ2020, JJ2021, JJ2023}. This implies that the coefficient $c_{j}$ of the position state $\ket{j}$ represents the probability amplitude at discrete position ${x_j}={j / 2^L}$, with normalisation $\sum_{j=0}^{2^L-1} |c_{j}|^2 = 1$. We note that the coordinate $x$ can be considered as a scaled 1D space, generated from the actual original domain $0 \le x' < R$ for a maximum distance $R$, with the scaled coordinate $x$ given in the domain of $0\le x = x'/|R| < 1$ with dimensionless absolute value $|R|$.

For quantum simulations with a first quantised approach, we define the translation operator $ \hat{A}_x$, known as the quantum adder \cite{Adder}. This shifts the position state such that $ \hat{A}_x \ket{j} = \ket{j+\delta x}$ \cite{JJ2020, JJ2021, JJ2023}, and so all the coefficients for the $L$-qubit ansatz state $\ket{\psi (x)}$ shift to the corresponding next grid point. As the grid separation is $\delta x = {1 / 2^L}$, the action is such that $\hat{A}_x \ket{\psi (x)} = \ket{\psi (x+\delta x)}$. We use $\hat{A}_x$ and its conjugate transposed operator $\hat{A}_x^{\dag}$, also known as the quantum subtractor. Note that we do not use the momentum operator itself with the translation operator, but the squared momentum operator $ \hat{p}^2  =  -\hbar^2 {d^2 / d x^2}$ can be approximated by the discretised expression for the second-order derivative operator in the finite difference method, up to the first-order precision, so that
\begin{eqnarray}
  { \hat{p}^{2}}  &=&    {-(mc \, {\cal L}_m)^2}  \left(\hat{\cal A}^{(1)} -2 \hat{\openone} \right) \; . \label{Expect-p-deriv-general}
\end{eqnarray}
Here, the dimensionless parameter is $ {\cal L}_m = {\lambdabar_m / \delta x}$ and we define $\hat{\cal A}^{(l)} = \left(\hat{A}_x \right)^l + \left(\hat{A}^{\dag}_x \right)^l$ generally, for any positive integer $l$ ($\hat{\openone}$: identity operator). Note that this approximation in Eq. (\ref{Expect-p-deriv-general}) does not diverge as $\delta x \rightarrow 0$ through an increase in qubit number $L$, because of $\hat{\cal A}^{(1)} \rightarrow 2 \hat{\openone} $ asymptotically as $\delta x \rightarrow 0$.

Furthermore, we should also point out that the parameters ($m$, $c$, $\hbar$ and $\delta x$) have to be selected with care to ensure that the perturbation criterion $\left< \hat{p}^2 \right> \ll m^2\,c^2$ is satisfied in any actual quantum simulation. For example, if $\hbar = m = c = 1$ and $\lambdabar_m = 1$ were to be selected, the relativistic effects would not be correctly treated in Eqs.~(\ref{Normal_HK01}) and (\ref{Kin-Energy01}). Here we utilise the translation operators ${\hat{A}_x}$ and ${\hat{A}^{\dag}_x}$ as unitary operators on quantum circuits, with the perspective of application to conventional quantum circuit platforms. However, in specific physical systems the operators could be implemented as non-unitary operators, if necessary for simulation with these systems \cite{JJ2016}.

\section{Relativistic kinetic energy} 
\subsection{Periodic boundary condition (PBC)}
We first consider the relativistic kinetic energy with periodic boundary conditions (PBC), which implies $\ket{\psi (0)}  = \ket{\psi (1)}$ for $0 \le x <1$ in the $x$ coordinate system. It is assumed that the translation operators $\hat{A}_x$ and $\hat{A}^{\dag}_x$ preserve the shape of the shifted wavefunction and its normalisation (i.e., $\hat{A}_x \ket{2^L-1} = \ket{0}$ and $\hat{A}^{\dag}_x \ket{0} = \ket{2^L-1}$), to maintain the property of unitary operators $\hat{A}^{\dag}_x \hat{A}_x = \hat{A}_x  \hat{A}^{\dag}_x = \hat{\openone}$. For PBC, the expectation values of even powers of momentum are given by 
\begin{eqnarray}
\left< \hat{p}^{2l} \right>_P &=& (-1)^l \, \hbar^{2l} \left< {d^{\, 2l} \over d x^{2l}} \right>_P ,  \nonumber \\
 & = &  \left( -(mc \, {\cal L}_m)^2 \right)^{l} \left<\,  \left( \hat{\cal A}^{(1)} - 2 \hat{\openone} \right)^l \, \right>\, .   \label{Expect-general-p} 
\end{eqnarray} 
For example, the first two terms with $l=1, 2$ (up to the fourth-order momentum operator) are given by combinations of $\left< \hat{\cal A}^{(l)} \right> $ for $l=1, 2$, such that
\begin{eqnarray}
\left< \hat{p}^{2} \right>_P &=& {-(mc \, {\cal L}_m)^2} \left(  \left< \hat{\cal A}^{(1)} \right> -2 \right) ,\label{Expect-general-p1} \\
\left< \hat{p}^{4} \right>_P &=& {(mc \, {\cal L}_m)^4} \left(  \left< \hat{\cal A}^{(2)} \right> - 4 \,  \left< \hat{\cal A}^{(1)} \right> + 6 \right).~~~~~\label{Expect-general-p2} 
\end{eqnarray}
Note that we need to compute only $\left< \hat{\cal A}^{(1)} \right> $ for the non-relativistic case, but also require $\left< \hat{\cal A}^{(2)} \right>$ for the first-order relativistic effect. 
Thus, we find that the generalised form of the relativistic kinetic energy with PBC is decomposed as combinations of the $\left< \hat{\cal A}^{(l)} \right>$, such that 
\begin{eqnarray}
\left< \hat{H}^{K }_{re} \right>_P  &=&  \beta_0 + \sum_{l=1}^{\infty} \beta_l \, \left< \hat{\cal A}^{(l)} \right> \,,
 \label{Energy02} 
\end{eqnarray} 
where the coefficients $\beta_l$ are given by
\begin{eqnarray} 
\beta_0 &=& 2 m c^2 \left( \alpha_1 \, {\cal L}_m^2 - 3 \alpha_2 \, {\cal L}_m^4 + ... \right),  \\
\beta_1 &=& m c^2 \left( - \alpha_1 \, {\cal L}_m^2 + 4 \alpha_2 \, {\cal L}_m^4 + ... \right), ~~~\\
\beta_2 &=& m c^2 \left(- \alpha_2 \, {\cal L}_m^4  + ... \right)\, .
\end{eqnarray}

We stress that the coefficients $\beta_l$ can be calculated beforehand, with appropriately chosen parameters of $m$, $c$, $\hbar$, and $\delta x$, for use in quantum simulations. Therefore, to estimate the relativistic kinetic energy terms up to the $l$-th order perturbation with PBC, we compute the expectation values up to $\left< \hat{\cal A}^{(l)} \right>$, to generate the higher-order squared momentum values up to $\left<\hat{p}^{2l} \right>_P$. 

\subsection{Dirichlet boundary condition (DBC)}
Dirichlet boundary conditions (DBC) allow more relaxed boundary conditions, as $\ket{\psi (0)}  \neq \ket{\psi (1)} $. A few recent works have studied the application of DBC for quantum computing approaches \cite{WES,Dieter}. In contrast to the PBC form of Eq.~(\ref{Expect-general-p}), for DBC the powers of the squared momentum take the form 
\begin{eqnarray}
 \left<  \hat{p}^{2l} \right>_D & = &   \left( -(mc \, {\cal L}_m)^2 \right)^{l}  \left<  \left(\hat{\cal A}^{(1)}  - 2 \hat{\openone}  - \hat{E}_0 \right)^l \right>,~~~ \label{Expect-general-k}
\end{eqnarray}
where a new operator is required, defined by 
\begin{eqnarray}
\hat{E}_0 = \ket{2^L -1}\bra{0} + \ket{0} \bra{2^L -1}\, . \label{E0-operator}
\end{eqnarray}

For $l=1$ in Eq.~(\ref{Expect-general-k}), the expectation value of the squared momentum relates to that for PBC of Eq. (\ref{Expect-general-p1}) through
\begin{eqnarray}
 \left<  \hat{p}^{2} \right>_D  & =&
 \left<  \hat{p}^{2} \right>_P + (mc \, {\cal L}_m)^2 \left< \hat{E}_0 \right> , ~~~\label{DB-Expect-p-deriv01}
\end{eqnarray}
with the expectation value of $\hat{E}_0$ given by
\begin{eqnarray}
\left< \hat{E}_0 \right> &=& \bra{\psi (x)} \hat{E}_0 \ket{\psi (x)} =  c^*_{2^L -1} \, c_{0} + c^*_{0} \, c_{2^L-1}\, . \label{E-expect01}
\end{eqnarray}
Therefore, the non-relativistic kinetic energy for DBC comprises the non-relativistic kinetic energy term for PBC $\left< \hat{H}^K_{nr} \right>_P$ plus the additional term $\left< \hat{E}_0\right>$, such that
\begin{eqnarray}
\left< \hat{H}^K_{nr} \right>_D  = \left< \hat{H}^K_{nr} \right>_P +  {1 \over 2} m\,c^2 ({\cal L}_m)^2 \left< \hat{E}_0 \right>\,.
\label{eq:DBC-nr01}
\end{eqnarray}
This implies that the data, from a non-relativistic kinetic energy example with the same ansatz state for PBC, can be reused to estimate the non-relativistic kinetic energy for DBC.

Moreover, the first-order relativistic effect in Eq.~(\ref{Normal_HK01}) is taken into account with $\left<\hat{p}^4 \right>$ for DBC, such that 
\begin{eqnarray}
&&  \left<  \hat{p}^{4} \right>_D  =  {(mc \, {\cal L}_m)^4} \left<  \left(\hat{ \cal A}^{(1)}   - 2 \openone - \hat{E}_0   \right)^2 \right>\,,  \label{Expect-p-deriv04}  \\
&&~~~=   \left<  \hat{p}^{4} \right>_P  + {(mc \, {\cal L}_m)^4} \Big( 4\left< \hat{E}_0 \right> - \sum_{k=1}^{2} \left< \hat{E}_i \right> + \left< \hat{E}^2_0 \right> \Big), \nonumber
\end{eqnarray}
for $\hat{E}_1 \equiv \hat{A}_{x} \hat{E}_0 + \hat{E}_0 \hat{A}_{x}^{\dag}$ and $\hat{E}_2 \equiv \hat{E}_0 \hat{A}_{x}+  \hat{A}_{x}^{\dag} \hat{E}_0$.

Since $\ket{\psi (x)} = \sum_{j=0}^{2^L-1} c_j \ket{j}$ in the domain ($0\le x < 1$) and 
\begin{eqnarray}
\hat{E}_1 &=&  \ket{1}\bra{2^L-1} + \ket{2^L-1} \bra{1}\, , \\
\hat{E}_2 &=&  \ket{0}\bra{2^L-2} + \ket{2^L-2} \bra{0} \, ,
\end{eqnarray}
the expectation values of $\hat{E}_1$, $\hat{E}_2$ and $\hat{E}_0^2 $ are given by
\begin{eqnarray}
\left< \hat{E}_1 \right> &=& c^{*}_{1}\, c_{2^L-1} + c^{*}_{2^L-1} \, c_{1} \,, \label{E1-operator} \\
\left< \hat{E}_2 \right> &=& c^{*}_{0}\, c_{2^L-2} + c^{*}_{2^L-2} \, c_{0} \,, \label{E2-operator} \\
\left< \hat{E}_0^2 \right> &=& |c_{0}|^2 + | c_{2^L-1}|^2 = {\cal P}_0 + {\cal P}_{2^L-1}\,, \label{E02-operator}~~~~~~
\end{eqnarray}
for the probabilities ${\cal P}_{j}$ of position state $\ket{j}$ in $\ket{\psi (x)}$. 

Therefore, the relativistic kinetic energy with the first-order perturbation for DBC $\left< \hat{H}^K_{re} \right>_D$ is computable from a combination of the relativistic kinetic energy term for PBC $\left< \hat{H}^K_{re} \right>_P$ in Eq.~(\ref{Energy02}) and additional expectation values $\left< \hat{F}_d \right>$, with specified coefficients $\gamma_d$, such that 
\begin{eqnarray}
\left< \hat{H}^K_{re} \right>_D = \left< \hat{H}^K_{re} \right>_P + \sum_{d=0}^3 \gamma_d \left< \hat{F}_d \right>\,.
\label{eq:DBC-Re01}
\end{eqnarray}
Here, $\hat{F}_d =  \{ \hat{E}_0, \,  \hat{E}_1, \,  \hat{E}_2, \,  \hat{E}^2_0\}$ 
and the specified coefficients are $\gamma_d = {mc^2} \{ \alpha_1 {\cal L}_m^2(1- {\cal L}_m^{2}), \, \alpha_2 {\cal L}_m^{4}, \, \alpha_2 {\cal L}_m^{4} , \, - \alpha_2 {\cal L}_m^{4} \} $ for $d=0,1,2,3$.  

\section{Quantum simulation algorithm}
We now provide the recipe for undertaking relativistic quantum simulations in quantum circuits, building on the formalism already presented. 
Since non-relativistic quantum simulations for PBC are more straightforward, because the methods for kinetic and potential energies have been investigated in Refs.~\cite{JJ2021, JJ2020}, we focus mainly on the implementation of the relativistic cases for PBC and DBC in quantum simulation, in particular the first-order perturbation terms of the kinetic energy for the two different boundary conditions.

\begin{figure} [t]
\centering
\includegraphics[width=0.59\textwidth,trim=1.5cm 11.5cm 0cm 1.5cm]{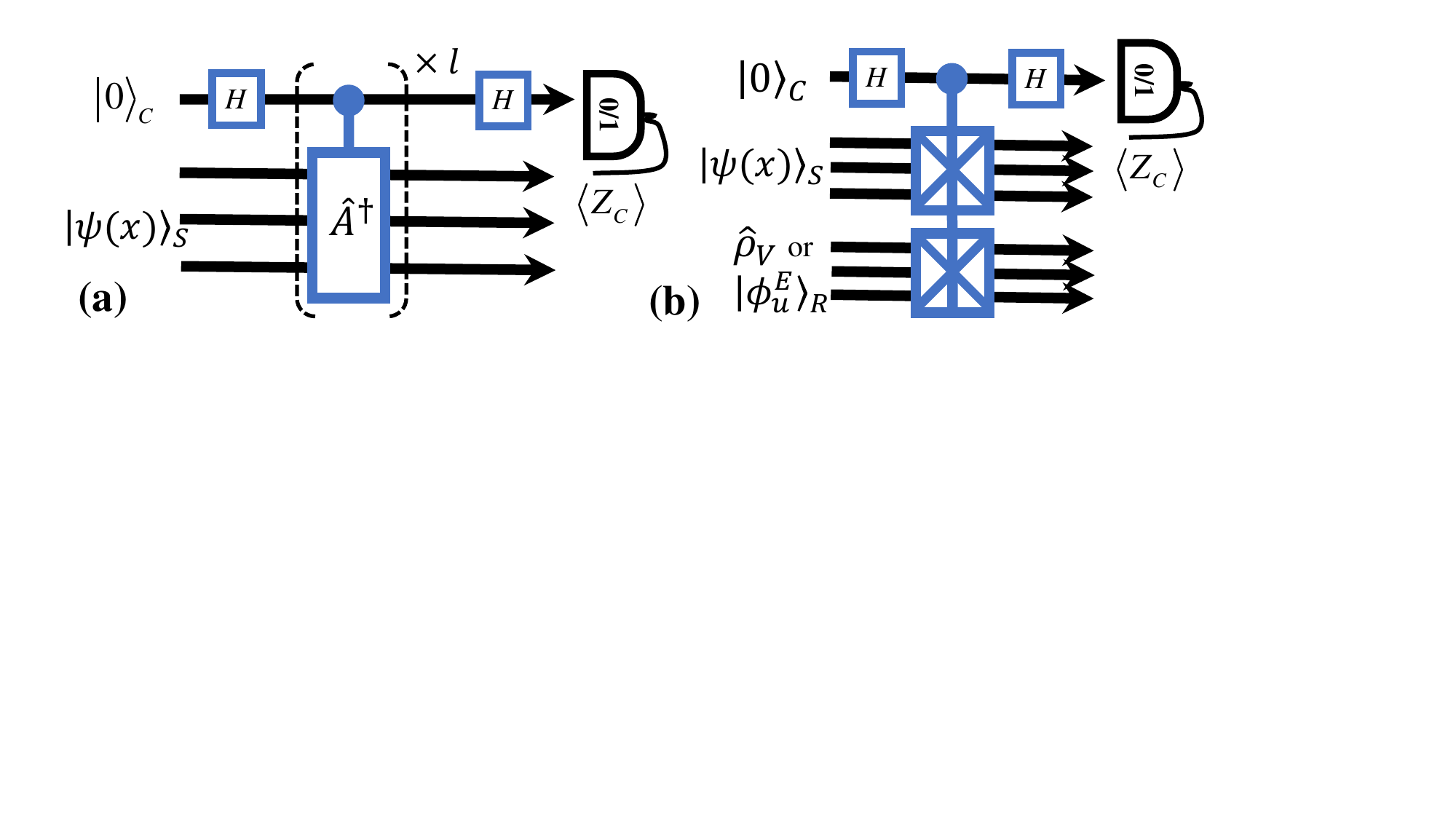}
\caption{Quantum circuits for (a) the expectation value $\Re \left[ (\hat{A}^{\dag}_x)^l \right]$ for the kinetic energy in PBC, (b) the potential energy $\langle \hat{V} \rangle$ and for preparing $L$-qubit state $\hat{\rho}_V$ for potential energy or $\ket{\phi^{E}_u}_{R}$ for kinetic energy in DBC.}
\label{fig:01}
\vspace{0.5cm}
\end{figure}

Fig.~\ref{fig:01} presents the designs for two quantum circuits to compute kinetic and potential energy terms for relativistic cases. The $L$-qubit spatial wavefunction $\ket{\psi (x)}$ is prepared in the system of qubits $S$ and a control qubit $\ket{0}_C$ in Fig.~\ref{fig:01}(a) \cite{JJ2023,JJ2020,JJ2021}. For simplicity, we restrict to the case that $c_j$ is a real-valued probability amplitude in the first-quantised wavefunction. After the initial Hadamard gate $H$ applied to $C$, we iteratively apply a controlled-$\hat{A}^{\dag}_x$ gate for $l$ times between $C$ and $S$. After a final Hadamard gate on $C$, we measure the control qubit $C$ in the $Z$-basis. The statistical result of $\left< {Z}_C \right>$ provides the desired expectation value $\Re \left[ \left< (\hat{A}^{\dag}_x)^l \right> \right] = \left< \hat{\cal A}^{(l)} \right>/2$, where $\Re [ \left< \cdot \right> ]$ is the real part of $\left< \cdot \right>$. It is also known that we can estimate the imaginary part of the expectation values by tuning the control qubit, if required \cite{JJ2023,JJ2020,JJ2021}. Since the quantities of $\left< \hat{p}^2 \right>_P$ and $\left< \hat{p}^4 \right>_P$ are given by $\Re \left[ \left< (\hat{A}^{\dag}_x)^l \right> \right]$ for $l=1, 2$ in Fig.~\ref{fig:01}(a), the kinetic energies for PBC are given by 
\begin{eqnarray}
\left< \hat{H}^K_{nr} \right>_P &=&  m c^2 {\cal L}_m^2 \left(  1- \Re \left[ \left<  \hat{A}^{\dag}_x \right> \right] \right), \\
\left< \hat{H}^{K }_{re} \right>_P  &=&  \beta_0 + 2 \sum_{l=1}^{2} \beta_l \, \Re \left[ \left<  \left( \hat{A}^{\dag}_x\right)^l \right> \right] \,.
 \label{eq:Energy04} 
\end{eqnarray}

For the kinetic energy terms in DBC, we need to compute the additional expectation values of $\hat{F}_d$ ($d=0, 1, 2, 3$) with the input reference state $\ket{\phi^E_u}_R$ shown in Fig.~\ref{fig:01}(b). For the three states $\ket{\phi^E_u}_R = \{ \ket{f}, \ket{g}, \ket{s} = (\ket{f} + \ket{g} )/\sqrt{2}\}$ individually prepared in reference qubits system $R$, each statistical outcome in the $Z$-basis measurements provides the probabilities of
\begin{eqnarray}
{\cal P}_f &=& \bra{\psi(x)} f \left> \right< f \ket{\psi(x)} =  |c_f|^2 \,, \label{expect_j} \\
{\cal P}_{g} &=& \bra{\psi(x)} g \left> \right< g \ket{\psi(x)} =  |c_{g}|^2\, , \label{expect_u} \\ 
{\cal P}_{s} &=& \bra{\psi(x)} s \left> \right< s \ket{\psi(x)} = |c_f + c_{g}|^2 , \nonumber \\
&=& {\cal P}_f + {\cal P}_{g} +  c^*_{f} \, c_{g} + c^*_{g} \, c_{f}, \label{expect_uj}
\end{eqnarray}
with $\ket{\psi (x)}= \sum_{j=0}^{2^L-1} c_j \ket{j}$. For example, if we perform the quantum circuit in Fig.~\ref{fig:01}(b) with $\ket{f} =\ket{0}$, $\ket{g}=\ket{2^L-1}$, and $\ket{s}=\left( \ket{0}+\ket{2^L-1}\right) /\sqrt{2}$ for $\ket{\phi^{E}_u}_{R}$, we find that $\left< \hat{F}_0 \right> = \left< \hat{E}_0 \right> = {\cal P}_{s} - {\cal P}_{f} - {\cal P}_{g}$, as given in Eq.~(\ref{E-expect01}).

\begin{table}[t]
\begin{center}
\begin{tabular}{| c | c |c |c| }
  \hline			
  Expect. value & $\ket{f}$ & $\ket{g}$ &  $\ket{s}$ \\ \hline
  $\left< \hat{E}_0 \right>$ or $\left< \hat{E}_0^2 \right>$ & ~$\ket{0}$~ & ~$\ket{2^L-1}$~ & ~$\left( \ket{0}+\ket{2^L-1}\right) /\sqrt{2}$ ~\\  \hline
 $\left< \hat{E}_1 \right>$  & ~$\ket{1}$~ & $ \ket{2^L-1}$ & $\left( \ket{1}+\ket{2^L-1}\right) /\sqrt{2}$ \\  \hline
$\left< \hat{E}_2 \right>$ & ~$\ket{0}$~ & $ \ket{2^L-2}$ & $\left( \ket{0}+\ket{2^L-2}\right) /\sqrt{2}$ \\  \hline
\end{tabular}
\caption{The set of states $\ket{\phi^{E}_u}_{R}$ for $\left< \hat{p}^{2} \right>_D$and $\left< \hat{p}^{4} \right>_D$.}
\label{table1}
\end{center}
\end{table}

Table \ref{table1} shows which reference states are injected in $\ket{\phi^E_u}_R $ to estimate the desired expectation values $\left< \hat{F}_d \right>$ in Eq.~(\ref{eq:DBC-Re01}) for DBC. $\left< \hat{F}_1 \right> = \left< \hat{E}_1 \right>$, and $\left< \hat{F}_2 \right> = \left< \hat{E}_2 \right>$ in Eqs.~(\ref{E1-operator}) and (\ref{E2-operator}) are evaluated with three different input states of $\ket{f}$, $\ket{g}$ and $\ket{s}$ in the second and third rows of Table \ref{table1}, whereas $\left< \hat{F}_3 \right> =\left< \hat{E}^2_0 \right>$ requires only $\ket{f}$ and $\ket{g}$ based on Eq.~(\ref{E02-operator}). Therefore, all the terms of the non-relativistic and relativistic kinetic energies in Eqs.~(\ref{eq:DBC-nr01}) and (\ref{eq:DBC-Re01}) for DBC are given by the combination of their corresponding kinetic energy terms for PBC and the additional terms $\left< \hat{F}_d \right>$ with fixed parameters $\gamma_d$ ($d=0,1,2,3$).

For the potential energy $\langle \hat{V} (x) \rangle$, we recycle the quantum circuit in Fig.~\ref{fig:01}(b) with a density matrix $\hat{\rho}_V$ to describe a desired potential $\hat{V} (x)$ \cite{JJ2020,JJ2021}. The shape of the potential in position space $x$ is represented in the diagonal density matrix $\hat{\rho}_V= \sum_{w = 0}^{2^L-1}   {\cal V}_w  \ket{w}\bra{w}$, because $\hat{V} (x) = {\cal S}\, \hat{\rho}_V (x) =  {\cal S}\,\sum_{w=0}^{2^L-1} {\cal V}_w \ket{w}\bra{w}$ where ${\cal S}$ represents the scale factor of the potential. Note that $ {\cal V}_w$ is pre-determined by the shape of the potential function in the $2^L$ grid points for $0 \le x <1$. 
Then, the potential energy with the system state $\ket{\psi (x)}$ is given by
\begin{eqnarray}
&& \langle \hat{V} (x) \rangle = {\cal S}\, \bra{\psi (x)}  \hat{\rho}_V (x)  \ket{\psi (x)} ={\cal S}\, \sum_{j= 0}^{2^L-1}   {\cal V}_j |c_{j}|^2,~~~~~~
\label{Expect_x-deriv03} 
\end{eqnarray}
for $tr \left( \,\hat{\rho}_V \right)  = \sum_{w=0}^{2^L-1} {\cal V}_w =1$. Therefore, the potential energy is given by the expectation values of the potential density matrix $\hat{\rho}_V$ with the probabilities of the system state ${\cal P}_j = |c_j|^2$ and the pre-computed coefficients ${\cal V}_j$ at position $x_j =j/2^L$.

In the light of the variational quantum simulation approach, the number of system qubits ($L$ qubits) is firstly determined and the system ansatz state is prepared in $\ket{\psi (\{ \bar{\kappa} \}, x)}_S = \sum_{j=0}^{2^{L}-1} c_{j}(\{ \bar{\kappa} \}) \ket{j}_S$ with the set of variational parameters $\{ \bar{\kappa} \}$. Dependent  upon the non-relativistic ($\mu = nr$) or the relativistic cases ($\mu = re$) under PBC or DBC (denoted by $\tau$), we can minimise the sum of the expectation values from the kinetic energy part $\left< \hat{H}^K_{\mu} \right>_\tau$ plus the potential energy $\langle \hat{V} \rangle_\tau$ with the same ansatz states $\ket{\psi (\{ \bar{\kappa} \}, x)}$. 
By tuning the variational parameters $\{ \bar{\kappa} \}$, we can estimate the total ground-state energy given by
\begin{eqnarray}
{\cal E}^{tot}_0 \approx  \min_{\{ \bar{\kappa} \}} \, \left( \left< \hat{H}^K_{\mu} \right>_\tau + \left< \hat{V} \right>_\tau \right) \, ,
\label{eq:TotalE01} 
\end{eqnarray}
for $\mu = nr~{\rm or}~re\,$ and $\tau = P~{\rm or}~D$.
It is very important to develop a method for effectively constructing ansatz states, which are well-suited to different problems in quantum simulation \cite{Chae-Yeun24}. Recently, an efficient method for preparing Gaussian states has been proposed in \cite{Gaussian-ansatz}.

\section{Conclusion and remarks}

In conclusion, this work presents a novel method for relativistic quantum simulation, using perturbation theory in a first quantisation framework. The method can estimate relativistic ground-state energies under both periodic and Dirichlet boundary conditions. By discretising the 1D wavefunction over a finite grid, the approach enables approximation of relativistic kinetic energies through perturbative expansions, incorporating higher-order momentum corrections. The quantum simulation circuits include ansatz states and controlled-SWAP or controlled-translation gates, to evaluate kinetic and potential energies in a relativistic quantum system.

Open research questions still exist with regard to achieving more accurate results from relativistic quantum simulations. 
First, increasing the number of system qubits is clearly equivalent to increasing the accuracy of the grid in the dimensionless position space, independent of the form of boundary conditions applied. The addition of a single qubit in the $L$-qubit ansatz state will clearly halve the position resolution $\delta x$. Nevertheless, the number of system qubits should be chosen with care. The resultant $\delta x$, along with the relativistic parameters $\hbar$, $m$, and $c$ used, must be such that the dimensionless parameter ${\cal L}_m$ is sufficiently small for perturbation theory to be useful. Otherwise, the contributions of the higher-order terms in the expansion will become dominant in Eq.~(\ref{Energy02}). 

Second, it is feasible to implement a higher-order Laplacian operator for $ {d^2 / d x^2}$, which could be related to the quantum approach of the partial differential equation solvers \cite{high-orderPrecision, high-orderPrecision2}. This approach  could generate contributions  equivalent to those in our higher-order perturbation terms. For example, the first-order perturbation terms contain $\left< \hat{\cal A}^{(1)} \right>$ and $\left< \hat{\cal A}^{(2)} \right>$, which are also utilised in the second-order derivative operator in the second-order precision. 
Along with specific relativistic examples, these and other open questions will be pursued in future work.

\section*{Acknowledgments}
This work was supported by the Ministry of Science, ICT and Future Planning (MSIP) by the National Research Foundation of Korea (RS-2024-00432214, RS-2025-03532992, RS-2023-00281456, and RS-2023-NR119931) and the Institute of Information and Communications Technology Planning and Evaluation grant funded by the Korean government (RS-2019-II190003, “Research and Development of Core Technologies for Programming, Running, Implementing and Validating of Fault-Tolerant Quantum Computing System”). This work is also supported by the Grant No. K25L5M2C2 at the Korea Institute of Science and Technology Information (KISTI).

\end{document}